\documentclass[preprint,12pt, a4paper]{elsarticle}
\usepackage{newunicodechar}
\newunicodechar{█}{\rule{0.8em}{1.6ex}}
\newunicodechar{║}{%
  \makebox[0pt][l]{\rule{0.07em}{1.6ex}}%
  \hspace{0.25em}%
  \rule{0.07em}{1.6ex}%
}

\usepackage{amssymb}
\usepackage{hyperref}
\usepackage{minted}
\usepackage[most]{tcolorbox}

\newcommand{\rev}[1]{\textcolor{black}{{#1}}}

\usepackage{listings}
\usepackage{xcolor}

\definecolor{yamlkey}{rgb}{0.0, 0.5, 0.0}      
\definecolor{yamlstring}{rgb}{0.7, 0.1, 0.1}   
\definecolor{yamlcomment}{rgb}{0.4, 0.4, 0.4}  

\lstdefinestyle{yamlstyle}{
  sensitive=true,
  keywords={true,false,True,False},
  keywordstyle=\color{black}\bfseries,          
  stringstyle=\color{yamlstring},               
  morestring=[b]',
  morestring=[b]",
  comment=[l]{\#},
  commentstyle=\color{yamlcomment}\itshape,
  identifierstyle=\color{yamlkey}\bfseries,    
  literate={:}{{\color{black}:}}1               
}

\definecolor{fortranturkis}{rgb}{0.0, 0.5, 0.5}  

\lstdefinestyle{fortranstyle}{
  language={[95]Fortran},                       
  sensitive=false,
  keywordstyle=\color{black}\bfseries,          
  commentstyle=\color{fortranturkis}\itshape,   
  stringstyle=\color{black},
  showstringspaces=false,
  alsoletter={_},
  literate={d0}{{\color{black}d0}}2
}

\definecolor{pykeyword}{rgb}{0.0, 0.2, 0.6}     
\definecolor{pystring}{rgb}{0.7, 0.1, 0.1}      
\definecolor{pycomment}{rgb}{0.4, 0.4, 0.4}     

\lstdefinestyle{pythonstyle}{
  language=Python,
  sensitive=true,
  keywordstyle=\color{pykeyword}\bfseries,
  stringstyle=\color{pystring},
  commentstyle=\color{pycomment}\itshape,
  showstringspaces=false,
  identifierstyle=\color{black}
}

\usepackage{ltablex}
\keepXColumns
\usepackage{longtable}

\journal{SoftwareX}

\begin{document}
\renewcommand{\labelenumii}{\arabic{enumi}.\arabic{enumii}}

\begin{frontmatter}
 


\title{MESAlab: A Python pipeline for MESA grid processing with integrated GYRE/MESA-RSP execution}


\author[label1,label2]{D. Tarczay-Nehéz}
\address[label1]{Konkoly Observatory, HUN-REN CSFK, MTA Centre of Excellence, Budapest, Konkoly Thege Miklós út 15-17, Hungary, tarczaynehez.dora@csfk.org}
\address[label2]{MTA–HUN-REN CSFK Lendület ”Momentum” Stellar Pulsation Research Group}

\begin{abstract}
Modules for Experiments in Stellar Astrophysics (\texttt{MESA}) is a widely used open‑source code for \rev{modeling} stellar evolution. 
Many applications of the code rely on large computational grids containing thousands of models, which can be time‑consuming to \rev{analyze}. 
The \texttt{mesalab} package was developed to automate this post‑processing. 
This Python‑based, modular pipeline supports the analysis of large \texttt{MESA} grid outputs and can identify selected evolutionary features, such as the blue loop phase of the evolution of classical Cepheids. 
The pipeline can automatically prepare input files for and launch \texttt{MESA‑RSP} and \texttt{GYRE} to investigate the pulsation properties associated with these evolutionary \rev{phases}.
The pipeline also includes tools for visualizing evolutionary tracks, including Hertzsprung–Russell diagrams and Gaia color–magnitude diagrams.

\end{abstract}

\begin{keyword}
astronomy \sep stellar evolution \sep Python pipeline \sep stellar pulsations \sep variable stars



\end{keyword}

\end{frontmatter}


\section*{Metadata}
\label{}

\begin{table}[!h]
\begin{tabular}{|l|p{6.5cm}|p{6.5cm}|}
\hline
\textbf{Nr.} & \textbf{Code metadata description} & \textbf{Metadata} \\
\hline
C1 & Current code version & v2.2.0 \\
\hline
C2 & Permanent link to code/repository used for this code version & \url{https://github.com/konkolyseismolab/mesalab} \\
\hline
C3  & Permanent link to Reproducible Capsule & \url{https://doi.org/10.5281/zenodo.20512505}\\
\hline
C4 & Legal Code License   & MIT License \\
\hline
C5 & Code versioning system used & git \\
\hline
C6 & Software code languages, tools, and services used & Python \\
\hline
C7 & Compilation requirements, operating environments \& dependencies & Runs on Python 3.9–3.11. Tested on Linux (Ubuntu 22.04) and macOS 12+. Required Python packages include numpy, pandas, matplotlib, scipy, astropy, PyYAML, tqdm, seaborn, numba, swifter, f90nml, isochrones, pygyre, and addict. Python 3.12+ is currently not supported due to dependency limitations. \\
\hline
C8 & If available Link to developer documentation/manual & \url{mesalab.readthedocs.io/en/latest/} \\
\hline
C9 & Support email for questions & tarczaynehez.dora@csfk.org\\
\hline
\end{tabular}
\caption{Code metadata (mandatory)}
\label{codeMetadata} 
\end{table}

\section{Motivation and significance}

Modeling and understanding stellar evolution is one of the most challenging fields of astronomy.
The Modules for Experiments in Stellar Astrophysics (\texttt{MESA}, \cite{Paxton2011,Paxton2013,Paxton2015,Paxton2018,Paxton2019,Jermynetal2023}) is a commonly used one-dimensional, versatile software instrument.
It offers various physical assumptions to adjust, such as mass, chemical composition, rotation, \rev{and} convective properties, allowing researchers to explore a wide range of evolutionary scenarios.
In many applications, the computational grids contain hundreds or thousands of models (e.g. \cite{Joyceetal2024,TarczayNehezetal2026}), making the post‑processing and analyzing output data highly time‑consuming.
To overcome this difficulty, the Python-based, open-source pipeline \texttt{mesalab}\cite{mesalab} was constructed.
The pipeline can be used primarily as a command‑line tool, configured via a simple YAML file that specifies the location of the \texttt{MESA} runs and the desired analysis steps and outputs. 
After setting the configuration file, the pipeline automatically performs blue loop detection on the selected grid directory and can generate Hertzsprung–Russell diagrams (HRDs) and combined Gaia color–magnitude diagrams (CMDs). 
It can also produce \texttt{GYRE} and \texttt{MESA-RSP inlists} for the selected models.
Furthermore, the individual modules of \texttt{mesalab} can also be imported and called directly as a Python package, allowing users to integrate the pipeline into custom analysis scripts or Jupyter notebooks.

A key motivation for developing the pipeline is the analysis of the so-called blue loop, a blue‑ward excursion on the HRD during \rev{the} core‑helium burning phase of classical Cepheids. 
The presence and extent of the blue loop are strongly dependent on the adopted input physics, making it a highly parameter‑sensitive phase of intermediate‑mass stellar evolution. \texttt{mesalab} automates the identification of models that enter the blue loop, generates HR diagrams for all evolutionary tracks, and produces a combined Gaia color-magnitude diagram and HRD for each detected loop. 
In addition, the pipeline can optionally run \texttt{GYRE} \cite{Townsendetal2013,Townsendetal2018,Goldstein2020,Sunetal2023} and \texttt{MESA-RSP} \cite{Paxton2019} calculations for asteroseismic analysis based on the generated \texttt{inlists}, with support for parallel execution to reduce computational time. 
This automated workflow significantly simplifies and accelerates the post‑processing of large \texttt{MESA} grids.

A recent study \cite{TarczayNehezetal2026} presents the first scientific application of the pipeline. 
Here, \texttt{mesalab} is used to map the parameter dependence of blue loops and to identify regions in the HRD and CMD where strange modes may occur \cite{Cox1980}. 
Strange modes are pulsation modes that arise from partial trapping in the outer stellar envelope, producing high‑overtone, low‑amplitude oscillations concentrated close to the stellar surface.
Building on the results of \cite{TarczayNehezetal2026}, a forthcoming paper applies \texttt{mesalab} to perform long‑term, nonlinear pulsation analyses on selected models. 

Beyond research applications, the pipeline has also proven useful in teaching: its modular structure and automated diagnostics make it suitable for introducing undergraduate students to stellar evolution modeling and asteroseismology.

\section{Software description}

\texttt{mesalab} is a Python-based analysis pipeline. 
It is designed to facilitate the analysis and post-processing of large \rev{grids} of outputs of stellar evolution models produced by the \texttt{MESA} software package.
\texttt{mesalab} is currently under active development, focusing on the automatic identification of Cepheid blue loop evolutionary phases, with planned extensions to detect additional evolutionary stages in future releases.

The package provides tools for scanning grids of \texttt{MESA} simulations in a given directory, extracting stellar parameters, identifying \rev{the} blue loop phase, and generating input data for asteroseismic analysis with \texttt{GYRE} and \texttt{MESA-RSP}.

\begin{figure}
    \centering
    \includegraphics[width=\linewidth]{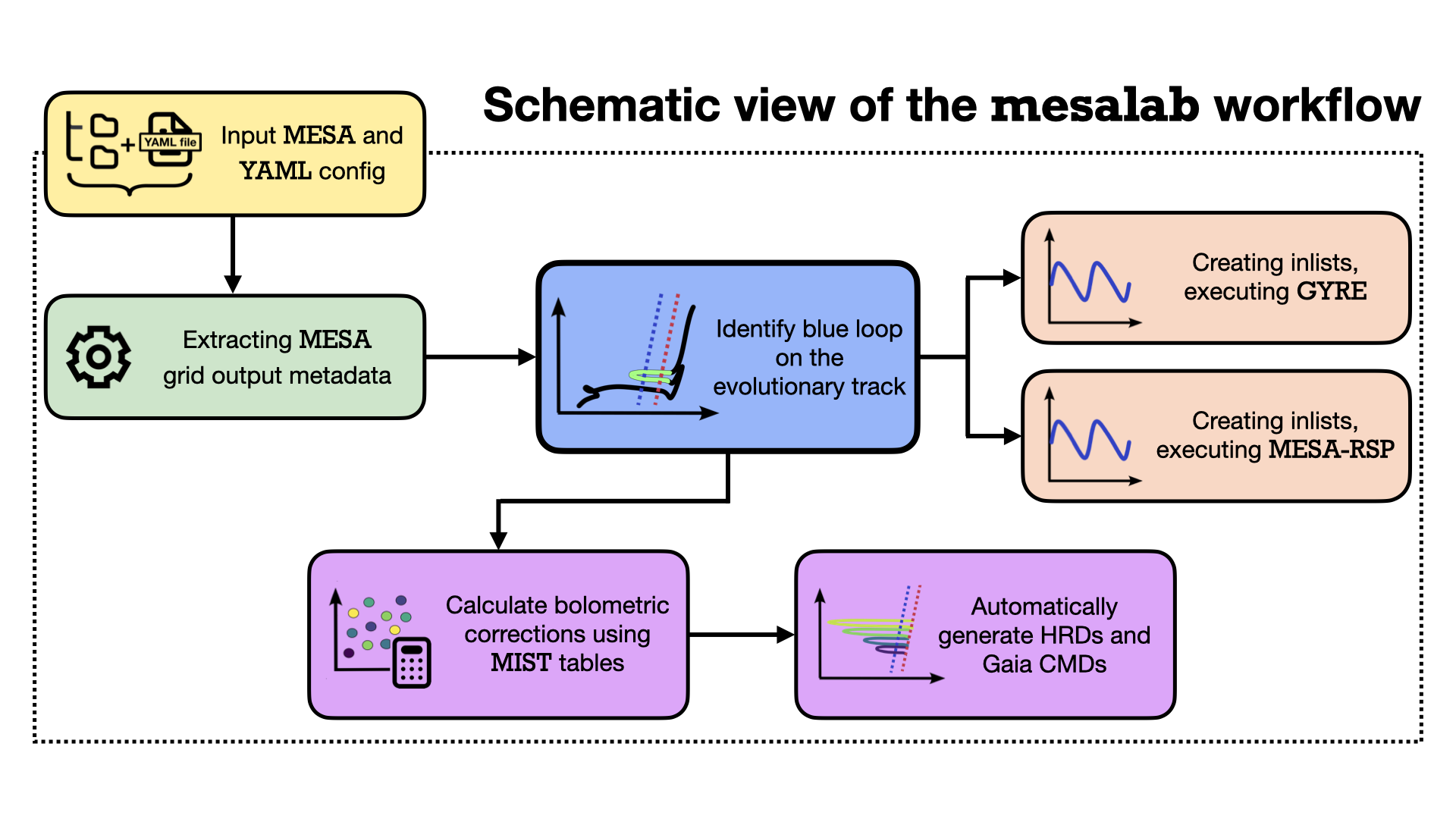}
    \caption{A schematic view of the \texttt{mesalab} pipeline.}
    \label{fig:flowchart}
\end{figure}

Figure~\ref{fig:flowchart} illustrates the overall workflow of the
\texttt{mesalab} pipeline.
Comprehensive documentation, including installation instructions, API references, and step‑by‑step tutorials, is available at \url{https://mesalab.readthedocs.io}.

\subsection{Software architecture}

The internal architecture of \texttt{mesalab} is organized into modular components that together form a configurable post‑processing pipeline. 
Each module is responsible for a distinct stage of the workflow, and the pipeline can be executed either through the command‑line interface or by importing the modules directly. 
The detailed API documentation is available in the online user guide of \texttt{mesalab} at \url{https://mesalab.readthedocs.io/en/latest/index.html}.

\subsubsection{Preprocessing layer}
\label{sec:preproc}
The pre-processing modules handle the discovery and loading of \texttt{MESA} simulations. 
A user‑defined root directory is scanned for subdirectories that contain both a Fortran \texttt{inlist} file and a \texttt{LOGS/history.data} file. 
The routines \texttt{extract\_params\_from\_inlist()} and \texttt{get\_mesa\_params\_from\_inlist()} parse the \texttt{inlist} and store the initial stellar parameters. 
Then the evolutionary tracks are loaded and stored by the \texttt{get\_data\_from\_history\_file()} routine from the \texttt{history.data} files. \rev{In this step, the general columns are needed from the \texttt{history.data} files, such as \texttt{log\_g}, \texttt{log\_Teff}, \texttt{log\_L}, \texttt{mass}, \texttt{star\_age}, \texttt{log\_R}, \texttt{center\_h1}, \texttt{center\_he4}, while parameters \texttt{initial\_mass} and
\texttt{initial\_Z} are retrieved from \texttt{inlist} files.}

\subsubsection{Data analysis layer}

The analysis modules operate on the processed evolutionary tracks. 
The blue loop analysis tool identifies those models that correspond to the blue loop phase (\texttt{analyse\_blue\_loop\_and\_instability()}).
\rev{This step identifies the end of the main sequence from the depletion of the central hydrogen abundance and subsequently determines the red giant branch tip as the minimum effective temperature reached before significant central helium depletion. The candidate blue loop phase is then restricted to the core helium-burning stage.}

\rev{\texttt{mesalab} then evaluates whether, during the core helium-burning phase, the evolutionary track crosses the instability strip from either the red or the blue side. These crossings are identified using point-in-polygon tests in the HR diagram with fixed, canonical instability strip boundaries defined on the red and blue edges. The canonical edges of the instability strip are consistent with the values adopted in Radek Smolec’s MESA tutorial\footnote{ See Radek Smolec’s tutorial in the \href{https://mesastar.org/summer-school-2023/}{MESA Summer School 2023}.}.}

The \texttt{perform\_mesa\_analysis()} function \rev{governs the analyzer} routine across the entire grid and prepares the data for the visualization and pulsation workflows.

\subsubsection{Visualization layer}

Diagnostic plots are generated by the visualization modules, which are activated based on the configuration settings. 
The \texttt{generate\_all\_hr\_diagrams()} routine produces Hertzsprung–Russell diagrams grouped by metallicity, while the heatmap utilities (\texttt{generate\_heatmaps\_and\_time\_diff\_csv()}) summarize instability strip crossing counts and phase durations across the grid. Bolometric corrections are calculated based on \texttt{MESA} Isochrones \& Stellar Tracks (MIST) Bolometric Correction Tables\footnote{\url{https://waps.cfa.harvard.edu/MIST/index.html}} \citep{MIST0,MIST} are applied when constructing Gaia color–magnitude diagrams. 
These tools operate on the filtered evolutionary snapshots selected during the analysis stage.

\subsubsection{Configuration and I/O layer}
The YAML configuration file controls the execution of the pipeline. 
The \texttt{parsing\_options()} function reads settings from default values, YAML files, environment variables, and command‑line arguments. 
The output manager creates the required output directory for all generated output, including summary CSVs, detailed blue loop files, and plot directories.

\subsubsection{Pulsation workflow layer for \texttt{MESA-RSP} and \texttt{GYRE}}

To facilitate and automate asteroseismic analysis, \texttt{mesalab} provides dedicated workflows for both \texttt{GYRE} and \texttt{MESA‑RSP}. 
\rev{Both \texttt{GYRE} and \texttt{MESA-RSP} are stellar pulsation codes based on different physical assumptions and numerical approaches. \texttt{GYRE} is a linear pulsation code applied to stellar models in hydrostatic equilibrium, capable of both adiabatic and non-adiabatic calculations. It is primarily used to determine mode frequencies and determine linear stability, including growth or damping rates. In contrast, \texttt{MESA-RSP} is a time-dependent nonlinear hydrodynamic code that follows the pulsation evolution of the star, allowing one to track the long-term behavior of stellar pulsations, e.g., the development of limit cycles and stable pulsation amplitudes. Within \texttt{mesalab}, the two approaches are used in a complementary way depending on whether linear stability diagnostics or nonlinear time-evolution modeling is required.}

During the earlier analysis stage, when the pipeline identifies the relevant evolutionary snapshots corresponding to the blue loop phase\rev{, i.e.,  those models that have entered the instability strip through the red edge}, the pipeline generates \texttt{GYRE} and \texttt{MESA‑RSP} \texttt{inlist} files based on a template user-defined template file. 
The pulsation workflow layer then executes these pre‑generated \texttt{inlists}.

The \texttt{GYRE} workflow is managed by \texttt{run\_gyre\_workflow()}, which executes \texttt{GYRE} either sequentially or in parallel, based on the configuration file. 
Individual runs are handled by \texttt{run\_single\_gyre\_model()}.

Similarly, the \texttt{MESA-RSP} is handled by the \texttt{run\_mesa\_rsp\_single()} for individual runs and by \texttt{run\_mesa\_rsp\_workflow()} for batch processing. 
Parallel execution is supported, and all outputs are organized into a dedicated \texttt{MESA-RSP} results directory.

\subsection{Software functionalities}

The main functionalities of \texttt{mesalab} include:

\begin{itemize}
\item automated scanning of user-defined directories containing grids of \texttt{MESA} stellar evolution models,
\item extraction of initial stellar parameters from \texttt{inlist} configuration files,
\item loading and preprocessing of evolutionary tracks from \texttt{history.data} files,
\item identification, selection, and characterization of blue loop phases and instability strip crossings,
\item generation of \texttt{inlist} files for both \texttt{GYRE} and \texttt{MESA-RSP},
\item optional execution of the asteroseismic workflows, including parallel batch processing,
\item production of diagnostic data such as Hertzsprung–Russell diagrams, Gaia color–magnitude diagrams.
\end{itemize}

\subsection{Sample code snippets}

The full usage of the pipeline, including end‑to‑end examples for grid analysis, blue loop detection, and running the \texttt{GYRE} and \texttt{MESA‑RSP} workflows, is demonstrated in the online tutorials.
These guides provide complete, executable examples and illustrate how to configure and run the pipeline both locally and in cloud environments such as Google Colab.

Comprehensive tutorials are available at:

\begin{itemize}
    \item Tutorial 1: \url{https://mesalab.readthedocs.io/en/latest/tutorial_1.html}
    \item Tutorial 2 (Google Colab example): \url{https://mesalab.readthedocs.io/en/latest/tutorial_2.html}
    \item Tutorial 3 (GYRE and RSP workflows): \url{https://mesalab.readthedocs.io/en/latest/tutorial_3.html}
\end{itemize}

These examples illustrate typical usage of the pipeline.

\section{Illustrative examples}

\subsection{Installation of the \texttt{mesalab} package}

The \texttt{mesalab} package requires a Python~3.9--3.11 environment, consistent
with the version constraints of its scientific dependencies (NumPy, SciPy,
Numba, Astropy). It can be installed using \texttt{pip}:

\begin{tcolorbox}[colback=black!5,colframe=black!50,title=Terminal]
\texttt{> pip install mesalab}
\end{tcolorbox}

\rev{The project is fully open-source and hosted on GitHub. For development or custom workflows, the package can also be cloned from the repository:
}

\begin{tcolorbox}[colback=black!5,colframe=black!50,title=Terminal]
\texttt{> git clone https://github.com/konkolyseismolab/mesalab.git}
\end{tcolorbox}

\rev{After cloning, users can install the package in editable mode, allowing direct modification of the codebase for custom analysis workflows and extensions.}

\begin{tcolorbox}[colback=black!5,colframe=black!50,title=Terminal]
\texttt{> pip install -e .}
\end{tcolorbox}

\subsection{Example configuration file for basic analysis}

\rev{The minimal input to the pipeline consists of a root directory containing a grid of \texttt{MESA} evolutionary models. Each model must be stored in a separate subdirectory and must contain an \texttt{inlist} file (see Section\,\ref{sec:preproc}) and a \texttt{LOGS/history.data} file.}

\rev{The pipeline automatically scans the directory structure, performs the full analysis, and writes all outputs to the specified output directory. The full analysis pipeline can be executed from the command
line using a YAML configuration file:}

\begin{tcolorbox}[colback=black!5,colframe=black!50,title=Terminal]
\begin{verbatim}
> mesalab --config example_MIST.yaml
\end{verbatim}
\end{tcolorbox}

A minimal example configuration file for analysing blue loop crossings and generating diagnostic plots may look as follows:

\begin{tcolorbox}[
    colback=black!5,
    colframe=black!50,
    title=\texttt{example\_MIST.yaml},
]
\begin{lstlisting}[style=yamlstyle, basicstyle=\ttfamily\small\color{black}, columns=fullflexible, keepspaces=true]
general_settings:
  input_dir: 'MIST_synthetic'
  output_dir: 'MIST_synthetic_output'
  inlist_name: "inlist_project"

blue_loop_analysis:
  analyze_blue_loop: True

plotting_settings:
  generate_hr_diagrams: "all"
  generate_blue_loop_plots_with_bc: True

gyre_workflow:
  run_gyre_workflow: False

rsp_workflow:
  run_rsp_workflow: False
\end{lstlisting}
\end{tcolorbox}

The \texttt{general\_settings} block specifies the input and output directories, \rev{and} the base \texttt{MESA} inlist name.  
The \texttt{blue\_loop\_analysis} section enables the identification and filtering of blueloop phases from the \texttt{MESA} evolutionary tracks.
The \texttt{plotting\_settings} control the generation of HR diagrams and the color-magnitude diagram.

The \texttt{gyre\_workflow} and \texttt{rsp\_workflow} sections are disabled in this minimal example and will be further discussed in Section \ref{sec:astero}.

Running the pipeline produces terminal output similar to:
\begin{tcolorbox}[colback=black!5,colframe=black!50,title=Terminal Output,breakable]
\begin{verbatim}
===========================================================
        mesalab CLI - Starting Analysis Workflow                         
                    Version: 2.2.0
===========================================================

===========================================================
       Starting MESA Analysis Workflow...
===========================================================

Performing MESA Run Analysis: 100%|██████████████|
                            21/21 [00:02<00:00,  7.59it/s]

===========================================================
       MESA Analysis Workflow Completed Successfully.
===========================================================

===========================================================
       Starting Plotting Workflow...
===========================================================

===========================================================
  Full Instability Strip Crossings Matrix (for Heatmap):
===========================================================
           2.1   4.6   6.1   7.6   9.1   10.6  12.1
initial_Z                                          
0.001420    0.0   2.0   2.0   2.0   2.0   2.0   0.0
0.014200    0.0   0.0   2.0   2.0   2.0   1.0   0.0
0.044904    0.0   0.0   0.0   0.0   0.0   0.0   0.0
===========================================================

Calculating BCs serially: 100%|█████████████████|
                           526/526 [00:05<00:00, 95.69it/s]

===========================================================
       Plotting Workflow Completed Successfully.
===========================================================

===========================================================
║         mesalab Workflow Finished Successfully!        ║
===========================================================
\end{verbatim}
\end{tcolorbox}

Upon successful completion of the workflow, \texttt{mesalab} generates a set of diagnostic plots in the output directory. 
These include Hertzsprung–Russell diagrams for all evolutionary tracks and Gaia color–magnitude diagrams for the detected blue loop phases. 
Examples of these automatically produced figures are shown below.

\begin{figure}
    \centering
    \includegraphics[width=\linewidth]{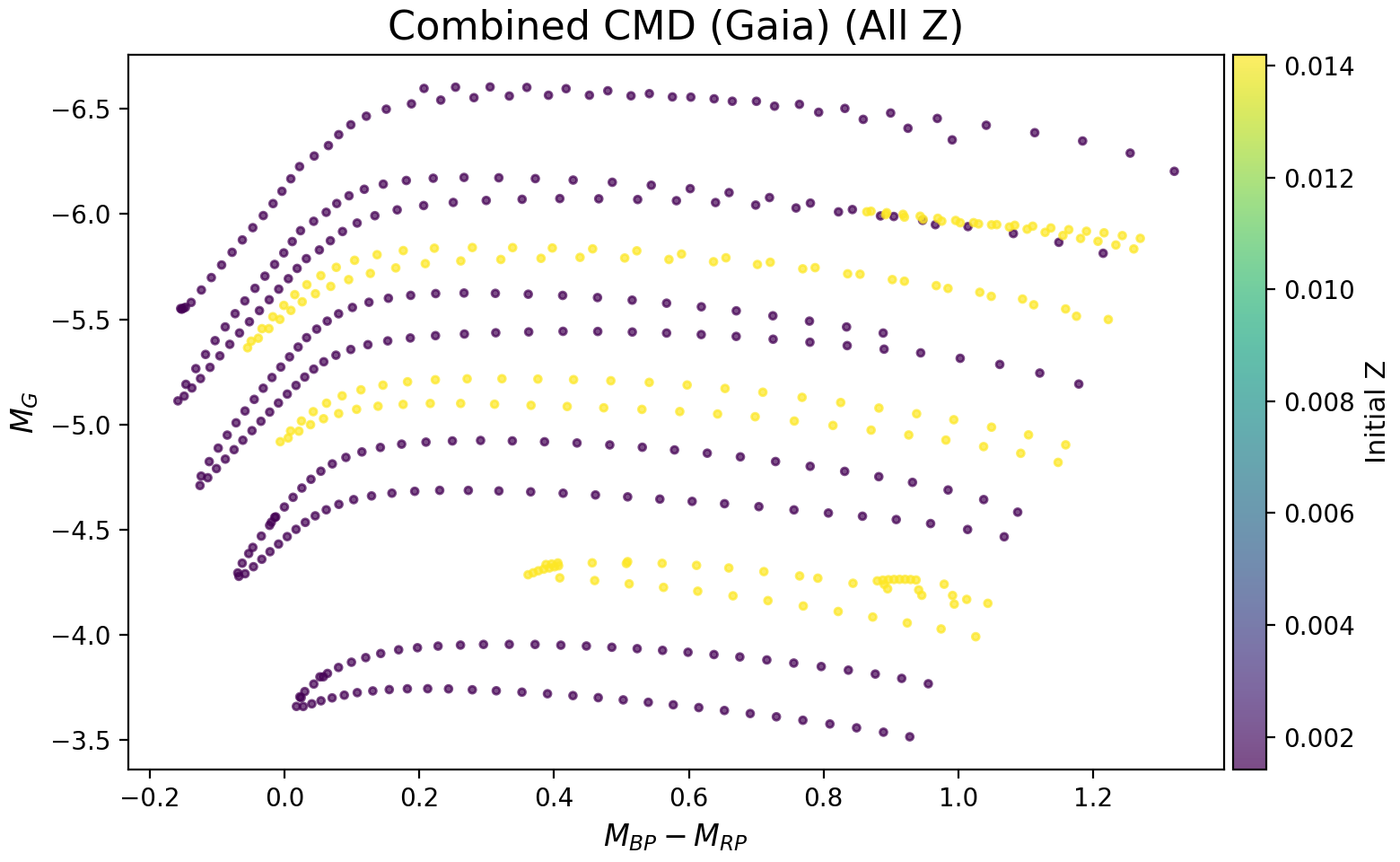}
    \includegraphics[width=\linewidth]{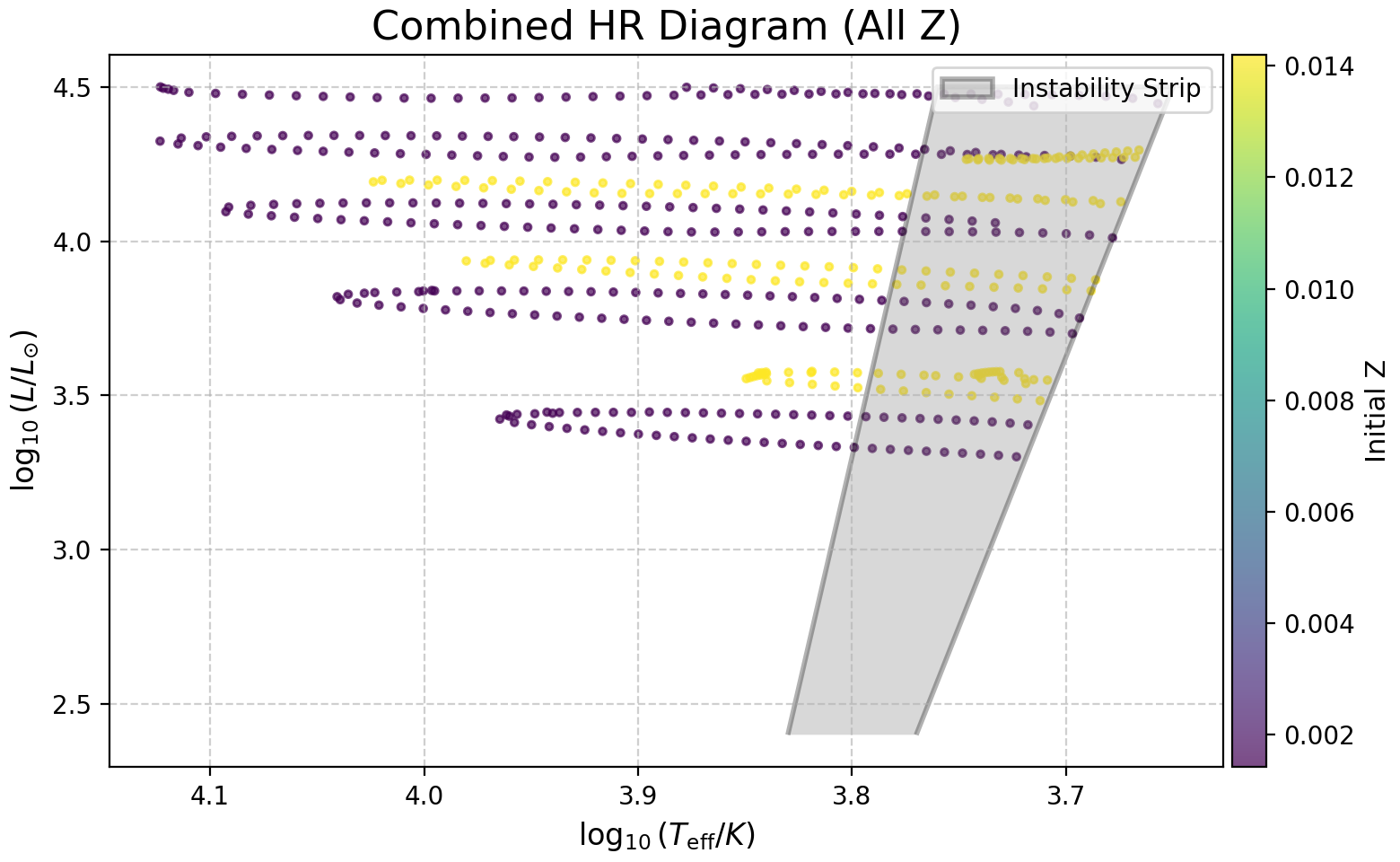}
    \caption{Example diagnostic plots generated by \texttt{mesalab}. 
    \textbf{Top:} Gaia color–magnitude diagram for models identified as entering the blue loop phase. 
    \textbf{Bottom:} Aggregated Hertzsprung–Russell diagram showing all evolutionary tracks in the grid, including models inside and outside the instability strip, with blue loop segments highlighted.}
    \label{fig:example}
\end{figure}

\subsection{Run the asteroseismic analysis modules}
\label{sec:astero}

\rev{The baseline execution shown in the example above demonstrates the standalone capabilities of the pipeline. The core preprocessing, data analysis (blue loop filtering, point-in-polygon instability strip validation), and visualization modules operate strictly within a standard Python environment. They require no local installation of \texttt{MESA} or \texttt{GYRE}, meaning users can fully analyze, filter, and plot pre-existing grid data logs on any standard desktop system.}

\rev{Conversely, e}nabling either the \texttt{GYRE} or \texttt{MESA--RSP} workflow requires a working installation of \texttt{MESA} and the \texttt{MESASDK}\rev{, as these are Python wrappers that actively run external binaries from \texttt{MESA} and \texttt{GYRE}.}
These components have been tested extensively with \texttt{MESA~r23.05.1} and \texttt{GYRE~7.0}; support for newer versions of \texttt{MESA} and \texttt{GYRE} will be extended in future releases.

\rev{To run these modules, the pipeline relies on user-provided template \texttt{MESA-RSP} or \texttt{GYRE inlist} files. The user must specify the exact file paths to these templates within the YAML configuration file.}

These modules can be enabled by setting
\texttt{run\_gyre\_workflow: True} or \texttt{run\_rsp\_workflow: True} in the
configuration file. For example:

\begin{tcolorbox}[
    colback=black!5,
    colframe=black!50,
    title=\texttt{example\_MIST.yaml} with \texttt{GYRE} and \texttt{MESA-RSP} enabled
]
\begin{lstlisting}[style=yamlstyle, basicstyle=\ttfamily\small\color{black}, columns=fullflexible, keepspaces=true]
gyre_workflow:
  run_gyre_workflow: True
  gyre_inlist_template_path: "../config/gyre.in"
  enable_gyre_parallel: True
  max_concurrent_gyre_runs: 8

rsp_workflow:
  run_rsp_workflow: True
  rsp_inlist_template_path: "../config/rsp.inlist_template"
  enable_rsp_parallel: True
  max_concurrent_rsp_runs: 4
\end{lstlisting}
\end{tcolorbox}

In this example, \texttt{run\_gyre\_workflow} and \texttt{run\_rsp\_workflow} enable the optional \texttt{GYRE} and \texttt{MESA--RSP} modules, respectively.
The \texttt{*\_inlist\_template\_path} fields specify the template input files from which \texttt{mesalab} generates run-specific inlists.
\rev{A simplified snippet of a \texttt{MESA-RSP} template configuration inside the \texttt{\&controls} namelist demonstrates this approach below:}

\begin{tcolorbox}[colback=black!5,colframe=black!50,title=\texttt{rsp.inlist\_template} snippet]
\begin{lstlisting}[style=fortranstyle, basicstyle=\ttfamily\small\color{black}, columns=fullflexible, keepspaces=true]
&controls
    ! ... standard MESA-RSP parameters ...
    
    ! Placeholders dynamically updated for each run:
    RSP_mass = 6d0
    RSP_Teff = 4892
    RSP_L = 4660
    RSP_X = 0.730d0
    RSP_Z = 0.003d0
    
    RSP_nmodes = 15
/
\end{lstlisting}
\end{tcolorbox}

\rev{Upon execution, \texttt{mesalab} locates these variables within the provided template file, modifies their values to match the exact properties of the selected evolutionary snapshot, and writes out a unique, runnable \texttt{inlist\_rsp} file into the corresponding output directory structure.}
The \texttt{enable\_*\_parallel} and \texttt{max\_concurrent\_*} parameters control parallel execution, allowing multiple \texttt{GYRE} or \texttt{RSP} calculations to run simultaneously.

The source code of \texttt{mesalab} is available at:
\url{https://github.com/konkolyseismolab/mesalab}

Detailed user documentation and tutorial notebooks are available at:
\url{https://mesalab.readthedocs.io}

The tutorials include complete, executable examples:
\begin{itemize}
    \item Tutorial 1: basic usage and grid scanning
    \item Tutorial 2: Google Colab example
    \item Tutorial 3: running \texttt{GYRE} and \texttt{MESA–RSP}
\end{itemize}

\subsection{Expected outputs and performance}
\label{sec:reproducibility}

\rev{To ensure the reproducibility of the pipeline, two distinct datasets are provided within the project repository and the Zenodo\footnote{\url{https://doi.org/10.5281/zenodo.20512505}} capsule. Both datasets can be processed through the entire pipeline:}

\begin{itemize}
    \item \rev{\texttt{MESA\_grid}: A compact $2\times2$ grid containing 4 actual, high-resolution \texttt{MESA} evolutionary tracks (4 and 5~M$_\odot$, $Z=0.0090$ and $0.0100$), spanning from the pre-main sequence to the post-blue loop phase.}
    \item \rev{\texttt{MIST\_synthetic}: This grid contains 21 evolutionary tracks derived from the MIST project (2.1 to 12.1~M$_\odot$). To ensure compatibility with \texttt{mesalab}, the original Equal-Evolutionary-Phase (EEP) files were pre-converted into quasi-history files with generated dummy \texttt{inlists} containing essential initial parameters ($M$, $Z$).}
\end{itemize}
\rev{The combined storage footprint of these test directories is approximately 728~MB.}

\rev{The total execution time depends strongly on the configurations. 
For basic analysis (with \texttt{run\_gyre\_workflow} and \texttt{run\_rsp\_workflow} set to \texttt{False}), the computational time is dominated by the multi-point interpolation of the bolometric correction tables. On a standard desktop computer, the sequential reading and processing of the 21 tracks in the \texttt{MIST\_synthetic} directory takes a few seconds, while the BC interpolation and diagnostic plot generation require a larger fraction of the execution time. However, the entire baseline data-filtering workflow for this dataset still finishes within the order of a minute.}

\rev{Conversely, if either of the \texttt{GYRE} or \texttt{MESA-RSP} modules is enabled, the total runtime will scale drastically depending on the number of selected models, the frequency resolution, and the hardware configuration. For instance, executing a full non-linear hydrodynamic \texttt{MESA-RSP} workflow pass over 373 selected stellar models from the test grid requires approximately 1~hour and 45~minutes using standard parallelization (\texttt{max\_concurrent\_rsp\_runs: 4}). Therefore, users should expect overall runtimes ranging from hours to days, or even several weeks, depending on the grid size and the specific pulsation settings (such as the number of modes, target periods, etc.).}

\rev{To avoid data loss during long-running grid executions, the pipeline saves its progress in real time. As a stellar model finishes, the pipeline instantly flushes its tracking metadata to disk. This ensures that even if a timeout or cluster failure occurs, the pipeline execution history is preserved up to the last processed model.}

\rev{Upon execution, the results are organized into the user-defined output directory (e.g., \texttt{MIST\_output}). Table~\ref{table:outputs} provides a detailed inventory of the generated directories, diagnostics, and the real-time \texttt{.json} workflow summary logs for both the \texttt{GYRE} and \texttt{MESA-RSP} modules.}

\footnotesize
\begin{longtable}{|p{0.5\linewidth}|l|p{0.4\linewidth}|}
\caption{Structure, directory taxonomy, and file inventory of the outputs generated by the \texttt{mesalab} pipeline modules, including the baseline execution and the optional asteroseismic extensions.} \label{table:outputs} \\
\hline
\textbf{Directory / File} & \textbf{Type} & \textbf{Description / Expected Content} \\
\hline
\endfirsthead
\hline
\textbf{Directory / File} & \textbf{Type} & \textbf{Description / Expected Content} \\
\hline
\endhead
\hline
\multicolumn{3}{|r|}{Continued on next page...} \\
\hline
\endfoot
\hline
\endlastfoot
\texttt{MIST\_output/} & Directory & The root output folder containing all results. \\
\hline
\texttt{$\hookrightarrow$ analysis\_results/} & Directory & Contains the aggregated grid-level outputs. \\
\texttt{\quad $\hookrightarrow$ crossing\_count\_grid.csv} & CSV table & Matrix of instability strip crossing counts. \\
\texttt{\quad $\hookrightarrow$ mesa\_grid\_time\_differences.csv} & CSV table & Log of precise stellar ages and calculated durations (in years) for both the blue loop phase and the instability strip crossings.\\
\texttt{\quad $\hookrightarrow$ processed\_runs\_overview.yaml} & YAML metadata & Overview of successfully processed grid subdirectories. \\
\texttt{\quad $\hookrightarrow$ summary\_results.csv} & CSV table & Master table summarizing the main physical features. \\
\hline
\texttt{$\hookrightarrow$ detail\_files/} & Directory & Contains detailed, model-by-model logs. \\
\texttt{\quad $\hookrightarrow$ detail\_z*.csv} & CSV table & Model-by-model evolutionary parameters grouped by metallicity, tracking stellar age, structure parameters ($\log T_{\mathrm{eff}}$, $\log L$, $\log g$), and the corresponding simulation directory paths.\\
\hline
\texttt{$\hookrightarrow$ plots/} & Directory & Contains all generated plots. \\
\texttt{\quad $\hookrightarrow$ CMD\_Gaia\_all\_blue\_loop\_data.png} & Plot & Combined Gaia color--magnitude diagram. \\
\texttt{\quad $\hookrightarrow$ HRD\_all\_blue\_loop\_data.png} & Plot & Global HRD including all tracks. \\
\texttt{\quad $\hookrightarrow$ LogL\_LogG\_all\_blue\_loop\_data.png} & Plot & $\log L$ vs $\log g$ diagram for structural analysis. \\
\texttt{\quad $\hookrightarrow$ HR\_diagram\_MESA\_grid\_z*.png} & Plots & Individual HR diagrams grouped by metallicity. \\
\texttt{\quad $\hookrightarrow$ mesa\_grid\_blue\_loop\_heatmap.png} & Plot & Visual heatmap of the crossing frequency matrix. \\
\hline
\texttt{$\hookrightarrow$ gyre\_output/} & Directory & Created when \texttt{GYRE} workflow is enabled. \\
\texttt{\quad $\hookrightarrow$ gyre\_workflow\_summary.json} & JSON summary & Real-time statistics file capturing tracking successful models and failed/timeout runs. \\
\texttt{\quad $\hookrightarrow$ run\_*/} & Directory & Model run directory matching the specific input folder name. \\
\texttt{\quad\quad $\hookrightarrow$ profile*/} & Directory & Target folder named after the filtered \texttt{profile} file. \\
\texttt{\quad\quad\quad $\hookrightarrow$ summary.h5} & Binary HDF5 & Overview of all calculated linear pulsation modes and eigenfrequencies. \\
\texttt{\quad\quad\quad $\hookrightarrow$ detail.l*.n*.TXT} & Plain text & Detail files containing eigenfunctions for a specific degree ($l$) and radial order ($n$), etc. \\
\hline
\texttt{$\hookrightarrow$ rsp\_outputs/} & Directory & Created when \texttt{MESA-RSP} workflow is enabled. \\
\texttt{\quad $\hookrightarrow$ rsp\_workflow\_summary.json} & JSON summary & Real-time statistics file capturing tracking successful models and failed/timeout runs. \\
\texttt{\quad $\hookrightarrow$ run\_*/} & Directory & Model run directory matching the specific input folder name. \\
\texttt{\quad\quad $\hookrightarrow$ model*/} & Directory & Target folder containing results for a specific filtered model number. \\
\texttt{\quad\quad\quad $\hookrightarrow$ inlist\_rsp} & Inlist file & The dynamically updated, unique runtime configuration file. \\
\texttt{\quad\quad\quad $\hookrightarrow$ rsp\_final\_*.mod} & Binary model & Saved state file of the final calculated nonlinear RSP model. \\
\texttt{\quad\quad\quad $\hookrightarrow$ photos/} & Directory & Contains structural snapshots and checkpoint data. \\
\texttt{\quad\quad\quad $\hookrightarrow$ LOGS/} & Directory & Subfolder containing standard \texttt{MESA-RSP} outputs. \\
\texttt{\quad\quad\quad\quad $\hookrightarrow$ LINA\_eigen*.data} & Plain text & Radial displacement eigenfunctions for the calculated modes. \\
\texttt{\quad\quad\quad\quad $\hookrightarrow$ LINA\_work*.data} & Plain text & Differential and cumulative work data for stability analysis. \\
\texttt{\quad\quad\quad\quad $\hookrightarrow$ LINA\_period\_growth.data} & Plain text & Growth rates and period listings for the selected radial modes. \\
\texttt{\quad\quad\quad\quad $\hookrightarrow$ history.data} & Plain text & Standard \texttt{MESA-RSP} time-series evolution log. \\
\texttt{\quad\quad\quad\quad $\hookrightarrow$ profile*.data} & Plain text & Final structure profile file. \\
\texttt{\quad\quad\quad\quad $\hookrightarrow$ profiles.index} & Plain text & Map index linking model numbers to specific profiles. \\
\hline
\end{longtable}

\subsection{Loading submodules to Pyhton}

\rev{In addition to the YAML command-line interface, \texttt{mesalab} can be imported as a standard Python module. The following example shows how to generate \texttt{MESA RSP} inlists from a \texttt{pandas} DataFrame and run the external \texttt{mesastar} binary file from a \texttt{Python} wrapper:}

\begin{tcolorbox}[colback=black!5,colframe=black!50,title=\texttt{example\_MIST.yaml} with \texttt{GYRE} and \texttt{MESA-RSP} enabled]
\begin{lstlisting}[style=pythonstyle, basicstyle=\ttfamily\small\color{black}, columns=fullflexible, keepspaces=true]
import pandas as pd
from addict import Dict
import mesalab.rsptools.rsp_inlist_generator as gen
from mesalab.rsptools.rsp_runner import run_mesa_rsp_workflow

# 1. Initialize DataFrame with required stellar parameters
data = {
    'initial_mass': [1.0], 
    'initial_Z': [0.014], 
    'model_number': [500],
    'log_Teff': [3.75], 
    'log_L': [1.5], 
    'initial_Y': [0.27],
    'run_dir_path': ['/path/to/mesa_run']
}
df = pd.DataFrame(data)

# 2. Generate the simulation inlists
inlists = gen.generate_mesa_rsp_inlists(
    detail_df=df, mesa_output_base_dir="/path/to/mesa_output",
    rsp_inlist_template_path="/path/to/rsp.inlist_template", 
    rsp_output_subdir="./rsp_out"
)

# 3. Configure and execute the parallel RSP workflow
config = Dict({
    'general_settings': {'mesa_binary_dir': 
    '/path_to_mesa_star_binary'},
    'rsp_workflow': {'enable_rsp_parallel': True, 
    'max_concurrent_rsp_runs': 4}
})
results = run_mesa_rsp_workflow(
    inlist_paths=inlists, 
    config_data=config, 
    rsp_output_subdir="./rsp_out"
)
\end{lstlisting}
\end{tcolorbox}

\rev{The complete API reference and detailed documentation for executing individual submodules are available at \url{https://mesalab.readthedocs.io/}.}

\section{Impact}

The \texttt{mesalab} pipeline simplifies the post‑processing of large grids of stellar evolution models by automating tasks that previously required custom scripts or extensive manual inspection. 
Instead of individually parsing hundreds or thousands of \texttt{MESA} runs, users can now identify instability strip crossings and blue loop phases of classical Cepheids, and select \rev{pulsationally} relevant evolutionary snapshots. 
A key advantage of the software is that the \texttt{GYRE} and \texttt{MESA‑RSP} modules can be run both within the pipeline and independently, enabling linear and nonlinear pulsation calculations to be performed consistently across entire evolutionary grids. 
This capability simplifies the process of systematic asteroseismic analysis of intermediate‑mass stars.

The pipeline has already been applied in published research \cite{TarczayNehezetal2026}, where it was used to identify the position of strange mode pulsating stars both on the Hertzsprung-Russell diagram and the Gaia color-magnitude diagram. 
The long‑term nonlinear modeling associated with this project is currently being developed in a follow‑up paper (in preparation). 

The software has also proven valuable in an educational context: its modular design and transparent workflow have enabled the involvement of undergraduate students in real research projects.

\rev{To demonstrate the quantitative reliability of the software, the (\texttt{example\_MESA\_base.yaml}) was used as a validation test. The pipeline automatically processes a test grid of 4 \texttt{MESA} evolutionary tracks, performing bolometric corrections across 373 individual snapshots. It robustly constructs the crossing matrix, identifying exactly 4 instability strip crossings within the $5.0\,M_\odot$ tracks. These automated results were verified via manual inspection of the evolutionary tracks, yielding identical results for all models.}

\section{Conclusions}

The \texttt{mesalab} pipeline offers a straightforward way to analyse \texttt{MESA} stellar evolution models and to prepare them for asteroseismic calculations with \texttt{GYRE} and \texttt{MESA–RSP}.
The individual steps of the analysis, such as selecting evolutionary phases, producing diagnostic plots, and generating pulsation input files, can be used together or separately, depending on the needs of the project.
Tutorials and example notebooks are available on the documentation website\footnote{\url{https://mesalab.readthedocs.io}}.

The future versions of \texttt{mesalab} will extend the automated phase‑identification capabilities besides the blue loop, allowing users to extract additional evolutionary stages of interest. 
This will further broaden the range of scientific questions that can be addressed with the pipeline.

\section*{Acknowledgements}
\label{}
This research was supported by the `SeismoLab' KKP-137523 \'Elvonal grant of the Hungarian Research, Development and Innovation Office (NKFIH), and by the LP2025-14/2025 Lendület grant of the Hungarian Academy of Sciences. 

During the preparation of this work, the author used Microsoft Copilot to assist with text editing and language refinement. After using this tool, the author reviewed and edited the content as needed and takes full responsibility for the content of the published article.





\end{document}